\documentclass[conference,letterpaper]{IEEEtran}
\IEEEoverridecommandlockouts % Without this \thanks{} don't appear
\usepackage[utf8]{inputenc} % The pack­age trans­lates var­i­ous stan­dard and other in­put en­cod­ings into a `LaTeX in­ter­nal lan­guage'
\usepackage[T1]{fontenc} % al­lows the user to se­lect font en­cod­ings, and for each en­cod­ing pro­vides an in­ter­face to `font-en­cod­ing-spe­cific' com­mands for each font
\usepackage[cmex10]{amsmath} % Use the [cmex10] option to ensure complicance
\usepackage{amssymb} % provides an extended symbol collection
\usepackage{mathtools} % optional corrections to amsmath, but recommended
\usepackage{amsfonts} % ex­tended set of fonts for use in math­e­mat­ics
\usepackage{accents} % A package for multiple accents in mathematics, with nice features concerning the creation of accents and placement of scripts

\usepackage{mleftright}\mleftright % defines variants \mleft and \mright of \left and \right, that make the delimiters act as \mathopen and \mathclose. These commands address spacing difficulties in subformulas.

\usepackage{multirow}
\usepackage[dvips]{graphicx} % for including graphics
\graphicspath{{./}}
\DeclareGraphicsExtensions{.eps}

\usepackage[dvipsnames]{xcolor} % provides easy driver-independent access to several kinds of colors, tints, shades, tones, and mixes of arbitrary colors; options: [xetex,dvipsnames]

\usepackage{bbm} % double-striked font for most symbols
\usepackage{xfrac} % diagonal fractions using \sfrac
\usepackage{cancel} % crossed-over equations; cancel-to-zero, etc. Options: [makeroom]
\usepackage{wrapfig} % typesetting a narrow float at the edge of the text, and making the text wrap around it.
\usepackage{tensor} % Al­lows the user to set ten­sor-style su­per- and sub­scripts with off­sets be­tween suc­ces­sive in­dices

\usepackage{theorem}
\usepackage{cite}

\usepackage{url} % The com­mand \url is a form of ver­ba­tim com­mand that al­lows line­breaks at cer­tain char­ac­ters or com­bi­na­tions of char­ac­ters, ac­cepts re­con­fig­u­ra­tion, and can usu­ally be used in the ar­gu­ment to an­other com­mand

\usepackage{psfrag} % Al­lows LaTeX construc­tions (equa­tions, pic­ture environments, etc.) to be pre­cisely superim­posed over En­cap­su­lated PostScript fig­ures

\usepackage[inline]{enumitem} % list customization; use [inline] to enable horizontal lists with *-ed environments

\usepackage{array} % extended  implementation  of  the  LaTeX array– and tabular–environments

\usepackage{aliascnt} % allows for a single counter in \autoref'ed theorems

\usepackage[draft]{hyperref} % cross-reference within document. Options: [hidelinks,linktocpage]

\usepackage{cleveref} % enhances LaTeX's cross-referencing features
\usepackage{xparse} % used for conditional macro definitions

\usepackage{ifthen} % Ifthen is a sep­a­rate pack­age within the LaTeX dis­tri­bu­tion; while it will al­ways be present in a LaTeX dis­tri­bu­tion, a \usep­a­ck­age com­mand is al­ways needed to load it

\usepackage{comment}

\crefname{section}{Section}{Sections}
\crefname{subsection}{Section}{Sections}
\crefname{equation}{Eq.}{Equations}
\crefname{enumi}{part}{parts}
\crefname{table}{Table}{Tables}

\newtheorem{theorem}{Theorem}
\crefname{theorem}{Theorem}{Theorems}

\newaliascnt{lemma}{theorem}
\newtheorem{lemma}[lemma]{Lemma}
\aliascntresetthe{lemma}
\crefname{lemma}{Lemma}{Lemmas}

\newaliascnt{definition}{theorem}
\newtheorem{definition}[definition]{Definition}
\aliascntresetthe{definition}
\crefname{definition}{Definition}{Definitions}

\newaliascnt{corollary}{theorem}
\newtheorem{corollary}[corollary]{Corollary}
\aliascntresetthe{corollary}
\crefname{corollary}{Corollary}{Corollarys}

\newaliascnt{claim}{theorem}

\aliascntresetthe{claim}
\crefname{claim}{Claim}{Claims}

\newaliascnt{conjecture}{theorem}

\aliascntresetthe{conjecture}
\crefname{conjecture}{Conjecture}{Conjectures}

\newaliascnt{question}{theorem}

\aliascntresetthe{question}
\crefname{question}{Question}{Questions}

\newaliascnt{oquestion}{theorem}

\aliascntresetthe{oquestion}
\crefname{oquestion}{Open Question}{Open Questions}

\theoremstyle{plain}
\theorembodyfont{\normalfont}

\newtheorem{cnstr}{Construction}%$\!$}

\newenvironment{construction}{\begin{cnstr}}{\hfill$\Box$\end{cnstr}}
\crefname{cnstr}{Construction}{Constructions}

\crefname{step}{Step}{Steps}

\crefname{regime}{Regime}{Regimes}

\newtheorem{myalgo}{Algorithm}%$\!$}

\crefname{myalgo}{Algorithm}{Algorithms}

\newcounter{enumrom}
\renewcommand{\theenumrom}{(\roman{enumrom})}

\makeatletter
\renewcommand{\@endtheorem}{\endtrivlist}
\makeatother

\makeatletter
\renewcommand{\thefigure}{{\@arabic\c@figure}}
\renewcommand{\fnum@figure}{{\bf Figure\,\thefigure}}
\makeatother

\renewcommand{\leq}{\leqslant}
\renewcommand{\geq}{\geqslant}

\newcommand{\bfc}{{\boldsymbol c}}

\newcommand{\bfu}{{\boldsymbol u}}
\newcommand{\bfv}{{\boldsymbol v}}
\newcommand{\bfw}{{\boldsymbol w}}
\newcommand{\bfx}{{\boldsymbol x}}
\newcommand{\bfy}{{\boldsymbol y}}

\newcommand{\cC}{\mathcal{C}}

\newcommand{\cF}{\mathcal{F}}

\newcommand{\cL}{\mathcal{L}}

\newcommand{\cR}{\mathcal{R}}

\newcommand{\cX}{\mathcal{X}}

\newcommand{\vcomment}[1]{{\color{violet}#1}}

\newcommand{\ecomment}[1]{{\color{Emerald}#1}}

\newcommand{\yy}[1]{{\footnotesize [\ecomment{#1}\;\;\vcomment{--Yoni}]}}

\DeclarePairedDelimiter\abs{\lvert}{\rvert}
\DeclarePairedDelimiter\ceilenv{\lceil}{\rceil}

\DeclarePairedDelimiter\parenv{\lparen}{\rparen}

\DeclarePairedDelimiter\bracenv{\lbrace}{\rbrace}
\DeclarePairedDelimiterX\mathset[2]{\lbrace}{\rbrace}{#1 : #2}
\DeclarePairedDelimiterX\inner[2]{\langle}{\rangle}{#1 \mathrel{},\mathrel{} #2}
\DeclarePairedDelimiterX\condparenv[2]{(}{)}{#1 \mathrel{}\delimsize\vert\mathrel{} #2}

\DeclareDocumentCommand\norm{ o m }{
    \IfNoValueTF{#1}
        {\left\Vert#2\right\Vert}
        {\left\Vert#2\right\Vert_{#1}}
}

\DeclareDocumentCommand\der{ o m o }{
    \IfNoValueTF{#1}
        {
            \IfNoValueTF{#3}
                {\frac{d}{d{#2}}}
                {\frac{d{#3}}{d{#2}}}
        }
        {\parenv*{\frac{d}{d{#2}}}^{#1}\IfNoValueTF{#3}{}{#3}}
}
\DeclareDocumentCommand\partder{ o m m }{
    \IfNoValueTF{#1}
        {\frac{\partial{#3}}{\partial{#2}}}
        {\frac{\partial^{#1}{#3}}{{\partial{#2}}^{#1}}}
}
\DeclareDocumentCommand\df{ o m o }{
    d\IfNoValueTF{#1}{}{^{#1}}{#2}\IfNoValueTF{#3}{}{_{#3}}
}

\newcommand{\tends}[1]{\underset{#1}{\longrightarrow}}

\newcommand{\deq}{\mathrel{\triangleq}}

\DeclareMathOperator{\rng}{rng}

\DeclareMathOperator{\red}{red}

\DeclareMathOperator{\rf}{\cR\cF}

\newcommand{\enc}[1]{\operatorname{Enc}_{\ref*{#1}}}

\begin{document}

\title{\textbf{Multi-strand Reconstruction from Substrings}}

\author{%
    \IEEEauthorblockN{\textbf{Yonatan  Yehezkeally}}
        \IEEEauthorblockA{%
            Institute for Communications Engineering\\
            Technical University of Munich\\
            Munich 80333, Germany \\
            yonatan.yehezkeally@tum.de} \and
    \IEEEauthorblockN{\textbf{Sagi Marcovich}}
        \IEEEauthorblockA{%
            Dept. of Computer Science\\
            Technion-Israel Institute of Technology\\
            Haifa 3200003, Israel \\
            sagimar@cs.technion.ac.il}  \and
    \IEEEauthorblockN{\textbf{Eitan Yaakobi}}
        \IEEEauthorblockA{%
            Dept. of Computer Science\\
            Technion-Israel Institute of Technology\\
            Haifa 3200003, Israel \\
            yaakobi@cs.technion.ac.il}
    \thanks{%
        This work was supported in part by the European Research Council (ERC) through the European Union's Horizon 2020 Research and Innovation Programme under Grant 801434. 
        Y. Yehezkeally was supported by the Alexander von Humboldt Foundation in cooperation with the Carl Friedrich von Siemens Foundation, under a postdoctoral research fellowship. S. Marcovich and E. Yaakobi were supported by the United States-Israel BSF grant no. 2018048.}%
}
\IEEEaftertitletext{\vspace{-2\baselineskip}}
\maketitle
\makeatother
\begin{abstract}
The problem of string reconstruction based on its substrings spectrum has received significant attention recently due to its  applicability to DNA data storage and sequencing. In contrast to previous works, we consider in this paper a setup of this problem where multiple strings are reconstructed together. Given a multiset $S$ of strings, all their substrings of some fixed length $\ell$, defined as the \emph{$\ell$-profile} of $S$, are received and the goal is to reconstruct all strings in $S$. A \emph{multi-strand $\ell$-reconstruction code} is a set of multisets such that every element $S$ can be reconstructed from its $\ell$-profile. Given the number of strings~$k$ and their length~$n$, we first find a lower bound on the value of $\ell$ necessary for existence of multi-strand $\ell$-reconstruction codes with non-vanishing asymptotic rate. We then present two constructions of such codes and show that their rates approach~$1$ for values of $\ell$ that asymptotically behave like the lower bound.
\end{abstract}

\section{Introduction} \label{sec:intro}
Reconstruction of strings refers to a large class of problems where the information about the string can only be provided in other forms than receiving it as one unit, even with possible errors. Examples for this set of problems are the \emph{$k$-deck problem}~\cite{DudSch2003,BenMeySchSmiSto91,Sco97} and the \emph{reconstruction from substring compositions problem}~\cite{AchDasMilOrlPan10,AchDasMilOrlPan15,MotBreTse13, MoRaTs13, GaMoRa16, BreBreTse13, SCT15, ShKaXiCoTs16}. Similar problems under this paradigm are the \emph{trace reconstruction problem}~\cite{BatKanKhaMcG2004} and the \emph{reconstruction problem} by Levenshtein~\cite{Lev01}, however in these setups the string is received as one unit multiple times with possible errors.

The problem of string reconstruction from its substring spectrum has received significant interest in the past decade and has been rigorously studied. For a length-$n$ string $\bfx$ and a positive integer $\ell$, its \emph{$\ell$-profile}, denoted by $\cL_\ell(\bfx)$, is the multiset of all its length-$\ell$ substrings. 
Then, the goal is to reconstruct the string $\bfx$ given only $\cL_\ell(\bfx)$. If a string can be uniquely reconstructed from its $\ell$-profile, then it is called \emph{$\ell$-reconstructible}. One of the main problems under this paradigm is to find the minimum value of $\ell$, as a function of $n$, which guarantees that the asymptotic rate of all $\ell$-reconstructible strings approaches 1. It was proved by Ukkonen~\cite{Ukk1992} that if all length-$\ell$ substrings of $\bfx$ are different from each other, then the string $\bfx$ is $(\ell+1)$-reconstructible. A string $\bfx$ that satisfies this constraint is referred to as \emph{$\ell$-repeat free}. Based upon this property, it was recently proved in~\cite{EliGabMedYaa21} that if $\ell=\lceil a\log(n)\rceil$ for some fixed value of $a>1$, then the asymptotic rate of all $\ell$-reconstructible strings approaches 1. The authors of~\cite{EliGabMedYaa21} also proposed two encoding schemes of $\ell$-repeat free strings; the first one uses a single redundancy symbol and supports $ \ell = 2\lceil \log (n)\rceil + 2$, while the second works for substrings of length 
$\ell = \ceilenv*{\log n} + \ceilenv*{2\log\log n} + 5$ 
and its asymptotic rate approaches 1.  Extensions of this problem to the setup where the $\ell$-profiles are not received error-free were also studied recently~\cite{MarYaa21, GabMil18, ChaChrEzeKia17, YehPol21}.

In this paper, we extend the problem of $\ell$-reconstructible strings to multisets of strings. This extension of the problem is motivated by DNA and polymer-based storage systems, since in both sequencing- and tandem mass spectrometry technologies it is typical that not a single string is read alone, but multiple strings simultaneously~\cite{salzberg_2010, Chin_2013, LoNiQuJoSiJa15, KhaPerBabNavSho18, GabPatMil21}. Thus, the $\ell$-profiles of all the strings in some multiset $S$ are read and the goal is to reconstruct all the strings in the multiset $S$. 
Assuming the multiset $S$ consists of $k$ length-$n$ strings, our first goal is to study the minimum value of $\ell$ as a function of $k$ and $n$ which guarantees the asymptotic rate of all $\ell$-reconstructible multisets approaches 1. We also present two efficient constructions of codes of $\ell$-reconstructible multisets where their asymptotic rate approaches 1.

The rest of this paper is organized as follows. \cref{sec:def} presents the definitions that will be used throughout the paper as well as the problem formulation of multi-strand $\ell$-reconstruction code. \cref{sec:lower bound} shows that if $\log(nk)-\ell = \omega_{nk}(1)$ then there do not exist positive-rate multi-strand $\ell$-reconstruction codes. In \cref{sec:codes}, two efficient constructions of multi-strand reconstruction codes are presented. We summarize and compare between the results of the constructions in the paper and show that, as a result, if $\ell \geq \log(nk) + 2\log\log(nk)+5$ then there exists a family of multi-strand reconstruction codes with asymptotic rate 1.\vspace{-0.5ex}

\section{Definitions and Preliminaries}\label{sec:def}\vspace{-0.5ex}

Let $\Sigma$ be a finite alphabet of size $q$, and denote some element 
$0\in\Sigma$. In our setting, information is stored in an 
unordered collection of $k$~strings of length~$n$ over $\Sigma$; it 
might be allowed for the same string to appear with multiplicity in 
the collection, which is encapsulated in the following formal 
definition. Let $\bracenv*{\bracenv*{a,a,b,\ldots}}$ denotes a 
multiset; i.e., elements are allowed to appear with multiplicity (for 
a multiset $S$, for convenience we let $\norm{S}$ denote the number of unique elements in $S$). Then 

\vspace{-1\baselineskip}
\begin{align*}
\cX_{n,k} &\deq \mathset*{S=\bracenv*{\bracenv*{\bfx_1,\ldots,\bfx_k}}}{\forall i: \bfx_i\in \Sigma^n}.
\end{align*}
Note that $\abs*{\cX_{n,k}} = \binom{k+q^n-1}{k}$.

For strings $\bfx,\bfy\in\Sigma^n$, we denote their concatenation by $\bfx\circ 
\bfy$. We say that $\bfv$ is a \emph{substring} of $\bfx$ if there exist 
strings $\bfu,\bfw$ (perhaps empty) such that $\bfx=\bfu\circ \bfv\circ \bfw$. If the length of $\bfv$ is $\ell$, we specifically say that $\bfv$ is an 
\emph{$\ell$-mer} of $\bfx$. 
Similarly, if $\bfx = \bfu\circ\bfv$ we say that $\bfu$ ($\bfv$) is a 
\emph{prefix} (\emph{suffix}, respectively) of $x$. An $\ell$-mer which is also a prefix is an \emph{$l$-prefix} (similarly, $\ell$-suffix). 
For a multiset $S\in \cX_{n,k}$, we let $\cL_\ell(S)$ denote its 
\emph{$\ell$-profile}, i.e., the multiset of all $\ell$-mers of all 
elements of $S$. For example, if $S = 
\bracenv*{\bracenv*{01010,00101,11101}}$ (which may be thought of as 
a multiset), 
\begin{align*}
    \cL_3(S) = 
    \bracenv*{\bracenv*{010,101,010,001,010,101,111,110,101}}.
\end{align*}
By abuse of notation, we let $\cL_\ell(\bfx)\deq 
\cL_\ell\parenv*{\bracenv*{\bracenv*{\bfx}}}$, for $\bfx\in\Sigma^n$.

For a window of length $\ell$, a code $\cC\subseteq \cX_{n,k}$ is said to 
be a \emph{multi-strand $\ell$-reconstruction code} if for all for all $S,S'\in \cC$ such that $S\neq S'$ it holds that 
$\cL_\ell(S)\neq\cL_\ell(S')$. 
Define 
\begin{align*}
A_{n,k,\ell} &\deq \mathset*{S\in \cX_{n,k}}{\cL_\ell(S)\ \text{is unique in}\ \cX_{n,k}}, \\
B_{n,k,\ell} &\deq \mathset*{\cL_\ell(S)}{S\in \cX_{n,k}}.
\end{align*}

The case of $k=1$ is of special interest; \cite{EliGabMedYaa21} 
introduced \emph{repeat-free} strings, which we denote herein for all $\ell <n $ by 
\begin{align*}
    \rf_\ell^n\deq \mathset*{\bfx\in\Sigma^n}{\norm{\cL_\ell(\bfx)} = n-\ell+1}.
\end{align*}
Observe that $\rf_\ell^n\subseteq A_{n,1,\ell+1}$~\cite{Ukk1992}. Moreover, an efficient algorithm reconstructs each $\bfx\in\rf_\ell^n$ from $\cL_{\ell+1}(\bfx)$ as follows. Given any $(\ell+1)$-mer $\bfv$ of $\bfx$, there exists a unique preceding $(\ell+1)$-mer $\bfu$ of $\bfx$ such that the $\ell$-suffix of $\bfu$ equals the $\ell$-prefix of $\bfv$ (unless $\bfv$ is a prefix of $\bfx$); similarly, a unique following $(\ell+1)$-mer. An extension of the same argument shows that 
if $S\in \cX_{n,k}$ satisfies $\norm{\cL_\ell(S)} = (n-\ell+1) k$, then $S$ is efficiently reconstructible from $\cL_{\ell+1}(S)$, and in particular, $S\in A_{n,k,\ell+1}$.

We note that $A_{n,k,\ell}$ is a multi-strand $\ell$-reconstruction code and that for any multi-strand $\ell$-reconstruction code $\cC\subseteq 
\cX_{n,k}$, $\abs*{\cC} \leq \abs*{B_{n,k,\ell}}$. 
For all $\cC\subseteq \cX_{n,k}$ (and, by abuse of notation, for 
$B_{n,k,\ell}$ as well) we denote the \emph{rate} and \emph{redundancy} of~$\cC$ by $R(\cC)\deq \frac{\log\abs*{\cC}}{\log\abs*{\cX_{n,k}}}$ and $\red(\cC)\deq \log\abs*{\cX_{n,k}}-\log\abs*{\cC}$, respectively. 
Throughout the paper, we use the base-$q$ logarithms where not otherwise indicated.

Finally, for two positive functions~$f,g$ of a common variable~$n$, we say that $f=o_n(g)$ if $\limsup_{n\to\infty}\frac{f(n)}{g(n)}=0$, $f=\Omega_n(g)$ if $\liminf_{n\to\infty}\frac{f(n)}{g(n)}>0$, and  $f=O_n(g)$ if $\limsup_{n\to\infty}\frac{f(n)}{g(n)}<\infty$. If clear from context, we omit the subscript from the aforementioned notations.

The main goal of this work is to find the minimum $\ell$, as a function of $n$ and $k$, such that the asymptotic rate of  $A_{n,k,\ell}$ and $B_{n,k,\ell}$ approaches 1. We will also be interested in efficient constructions of multi-strand $\ell$-reconstruction codes with asymptotic rate 1 while the value of $\ell$ will be close to the minimum value that accomplishes this rate result.

\section{A Lower Bound on $\ell$ for Codes with Positive Rate}\label{sec:lower bound}
We begin by analyzing our channel, $\cX_{n,k}$. Throughout this work, 
we assume in asymptotic analysis that as $n$ grows, $\alpha\deq 
\limsup\frac{\log(k)}{n} < 1$.

\begin{lemma}\label{lem:channel}
$\log\abs*{X_{n,k}} = n k - k \log(k/e) + o(k)$. 
\end{lemma}
\begin{IEEEproof}
We note that 
\begin{align*}
\frac{q^{nk}}{k!} 
\leq \abs*{\cX_{n,k}} 
&= \frac{q^{nk}}{k!} \prod_{j=0}^{k-1}\parenv*{1 + \frac{j}{q^n}} \\
&\leq q^{nk} \parenv*{\frac{e}{k}}^k 
\parenv*{\frac{1}{k}\sum_{j=0}^{k-1}\parenv*{1 + \frac{j}{q^n}}}^k \\
&\leq q^{nk} \parenv*{\frac{e}{k} + \frac{e}{2q^n}}^k.
\end{align*}

Recalling that $\log(k!)\leq \log\parenv*{e \sqrt{k} (k/e)^k} = 
k\log(k/e) + O(\log(k))$, and observing for 
$\alpha<1$ that $\frac{e}{k} + \frac{e}{2q^n} = \frac{e}{k} 
\parenv*{1+O(q^{-(1-\alpha) n})}$, the proof is concluded.
\end{IEEEproof}

We can now observe a lower bound on the required window length $\ell$ for multi-strand $\ell$-reconstruction codes to have positive rate and in particular rate approaching 1.
\begin{lemma}\label{lem:zero-rate}
Let $f:(0,\infty)\to(0,\infty)$ be any function satisfying $f(x) < 
\log(x)$ and $f(x)\tends{x\to\infty}\infty$. Let $\ell\leq\log(n k) - 
f(n k)$. Then,
\begin{align*}
    R(A_{n,k,\ell}) \leq R(B_{n,k,\ell}) = o_{n k}(1).
\end{align*}
\end{lemma}
\begin{IEEEproof}
We follow \cite{ChaChrEzeKia17}, by defining \emph{profile vectors}, as follows. For every $S\in \cX_{n,k}$ and $\bfv\in\Sigma^\ell$, let $f_S(\bfv)$ count the number of times $\bfv$ appears in $\cL_\ell(S)$. 
Clearly, if $S,S'\in \cX_{n,k}$ satisfy $\cL(S)\neq\cL(S')$, then $f_S\neq f_{S'}$. Since $\sum_{v\in\Sigma^\ell} f_S(\bfv) = k(n-\ell+1)$, this implies that $\abs*{B_{n,k,\ell}} \leq \binom{k(n-\ell+1) + q^\ell-1}{q^\ell-1}$. Observe that $\frac{q^\ell}{n k}\tends{n k\to\infty} 0$; then, based on $\binom{a}{b}\leq 2^{a H_2(b/a)}$, where $H_2$ is the binary entropy function (see, e.g., \cite[Ch.10,~Sec.11,~Lem.7]{MacSlo78}), it is possible to derive that $\log\abs*{B_{n,k,\ell}} = o(n k)$. 
Recalling \cref{lem:channel}, we have that $\log\abs*{\cX_{n,k}} \geq\parenv*{n-\log(k)} k = \Omega(n k)$, and thus the proposition is proven. 
\end{IEEEproof}

\section{Constructions of Multi-strand Reconstruction Codes}\label{sec:codes}
In contrast to \cref{lem:zero-rate}, we show in this section that if $\ell \geq \log(nk) + 2\log\log(nk) + 5$ as $nk$ grows, then $R(B_{n,k,\ell}) \geq R(A_{n,k,\ell}) = 1 - o_{nk}(1)$. 
We shall do so by presenting two explicit constructions of multi-strand $\ell$-reconstruction codes with efficient encoders and decoders.

Our constructions will be generic in the sense that they will apply 
any repeat-free encoding/decoding algorithm, and more specifically 
the ones from~\cite{EliGabMedYaa21}. Since the algorithms 
from~\cite{EliGabMedYaa21} use another component of run-length limited 
constrained (RLL) codes (for discussion of their use there, see the 
proof of \cref{thm:red-naive}), we first begin with an examination of 
known encoders for the well-studied $(0,M-1)_q$-RLL constraint (see, 
e.g., \cite[Sec.~1.2]{MarRotSie01}). The formal definition of these 
codes is given as follows.
\begin{definition}
Let $Z(N,M)$ denote the set of length-$N$ strings over $\Sigma$ 
containing no zero-run of length $M$.
\end{definition}
While most previous works studied the case where $M$ is fixed with respect to $N$,~\cite{LevYaa19} studied this constraint when $M=\log(N)+O(1)$, and showed that $\red\parenv*{Z(N,M)} = \Theta\parenv*{N/q^M}$. Even though we could use the results from~\cite{LevYaa19} for the derivation of the results in our paper, we next show how they can yet be improved. These improvements will be beneficial when deriving the parameters of the multi-strand reconstruction codes generated by our two constructions in this section. We start with the following lemma on the redundancy of the set $Z(N,M)$.
\begin{lemma}\label{lem:rll-red} 
$\red\parenv*{Z(N,M)}\leq \frac{q-1}{q} (1+o_M(1)) \frac{N}{q^M}$. 
\end{lemma}
\begin{IEEEproof}
It is well-known (see, e.g., for the binary case \cite[Exm.~3.3]{MarRotSie01}), that for a fixed $M>0$, $\lim \frac{1}{N}\log\abs*{Z(N,M)} = 
\log(\lambda)$, where $\lambda$ is the unique real root strictly greater than 
$1$ of the polynomial $p(x) = x^{M+1}-qx^M+q-1$. Since $\log\abs*{Z(N,M)}$ is 
sub-additive in $N$, by Fekete's lemma (as was observed in 
\cite{EliGabMedYaa21}) $\lim \frac{1}{N} \log\abs*{Z(N,M)} = 
\inf \frac{1}{N} \log\abs*{Z(N,M)}$ for any fixed $M$. Hence, in particular, 
for all $N$ it may be observed $\log\abs*{Z(N,M)} \geq N \log(\lambda)$. 
Equivalently, $\red\parenv*{Z(N,M)}\leq N \parenv*{1-\log(\lambda)}$. Based on 
\cite[App.]{JaiFarSchBru17a}, $1-\log(\lambda) \leq \frac{q-1}{q} \cdot 
\frac{1+o_M(1)}{q^M}$, which completes the proof.
\end{IEEEproof}

The authors of~\cite{LevYaa19} also presented an algorithm with efficient encoder/decoder pair from $\Sigma^{N'}$ into $Z(N,M)$ (see Alg.~1 and the discussion at the end of Section III in~\cite{LevYaa19}). It analyzed the case $q=2$, and the resulting encoder requires $N-N' = 2\ceilenv*{N/q^{M-1}}$ redundant symbols, which is the optimal order of magnitude. We can slightly improve upon this algorithms, when $q>2$. 
\begin{lemma}
If $q>2$, an efficient encoder/decoder pair into $Z(N,M)$ exists, 
requiring $\ceilenv*{N/\parenv*{q^{M-1}(q-2)+M-1}}$ redundant symbols. 
\end{lemma}
\begin{IEEEproof}
The concept of the encoder is similar to \cite[Alg.~1]{LevYaa19}. 
First, the information string $\bfx\in\Sigma^{N'}$ is divided into blocks of length $n$, to be determined later. Then, in each block:
\begin{enumerate}
    \item Append a $1$.
    \item From left to right, search for zero-runs of length $M$; if one is encountered, remove it, and append the index of its incidence to the block using $M$ symbols, such that the last symbol is restricted not be be either $\bracenv*{0,1}$.
    \item Continue, until no further zero-runs of length $M$ exist.
\end{enumerate}
Note that this process concludes in finite time (as in each iteration of step 2 it must advance in finite time, and the string length is preserved). Further, with the given restriction, $M$ symbols may index a total of $q^{M-1}(q-2)$ positions for the beginning of the zero $M$-mer. It is therefore required to set $n=q^{M-1}(q-2)+M-1$.

Also observe that a possible decoder can use the last symbol to indicate whether a zero-run of length $M$ was removed and indexed (which it can then inject in the correct place, discarding the index), or if the process is concluded (in which case the $1$ suffix should also be discarded).

Next, since every encoded block ends with a nonzero symbol, these blocks can be concatenated without violating the constraint. Observe, then, that a single redundant symbol is added per block, hence the claimed overall redundancy.

Finally, note that both encoder and decoder operate in polynomial time in the input length.
\end{IEEEproof}

We are now ready to present two distinct constructions for 
multi-strand reconstruction codes. For convenience, we assume all 
quantities to have integer values; a straightforward adjustment of 
the described methods applies for all values.

\subsection{Construction A}
Our first construction of multi-strand reconstruction codes is next presented.
\begin{construction}\label{cnst:naive}
Let $\bfx\in\Sigma^m$ be an arbitrary information string, and encode it 
into an $\ell'$-repeat-free string $\bfc=E(\bfx)\in\Sigma^{n'}$ using any known repeat-free encoder. Take $\bfc_1,\ldots,\bfc_k\in\Sigma^{n'/k}$ such that $\bfc = \bfc_1\circ \bfc_2\circ \cdots\circ \bfc_k$. Let $f(i)\in\Sigma^{\log(k)}$ be a $q$-ary expansion of $i\in[1,k]$. Denote $\widetilde{\bfc}_i \deq f(i)\circ \bfc_i$, and let $n\deq n'/k + \log(k)$. Then, \vspace{-0.5ex}
\begin{align*}
\enc{cnst:naive}(\bfx) \deq 
\bracenv*{\mathset*{\widetilde{\bfc}_i}{i=1,\ldots,k}}\in \cX_{n,k}.\vspace{-2ex}
\end{align*}\vspace{-2ex}
\end{construction}
The decoding success of \cref{cnst:naive} follows from the next lemma. 
\begin{lemma}
For all $\bfx\in\Sigma^m$, it holds that $\enc{cnst:naive}(\bfx)\in 
A_{n,k,\ell+1}$, where $\ell = \ell'+\log(k)$.
\end{lemma}
\begin{IEEEproof}
Note that $\bfc = \bfc_1\circ \bfc_2\circ \cdots\circ \bfc_k \in \rf_{\ell'}^{n'}$ and thus $\norm{\cL_{\ell'}\parenv*{\bracenv*{\mathset*{\bfc_i}{i=1,\ldots,k}}}} = (n'-\ell'+1) k$. It follows that $\bracenv*{\mathset*{\bfc_i}{i=1,\ldots,k}} \in A_{n',k,\ell'+1}$.

Now, let $\bfu,\bfv$ be $(\ell+1)$-mers of $\widetilde{\bfc}_i, 
\widetilde{\bfc}_j$ respectively; note that the $(\ell'+1)$-suffixes 
of $\bfu,\bfv$ are $(\ell'+1)$-mers of $\bfc_i, \bfc_j$ respectively, and hence 
if $\bfu=\bfv$ then $i=j$ and the positions of both $(\ell'+1)$-mers agree. 
It follows that the positions of $\bfu,\bfv$ agree as well, and the claim 
follows.
\end{IEEEproof}

Recall, then, that given $\cL_{\ell+1}(\enc{cnst:naive}(\bfx))$, an efficient algorithm can easily produce the set of strings $\widetilde{\bfc}_1,\widetilde{\bfc}_2,\ldots,\widetilde{\bfc}_k$ by simple stitching based on identical prefixes and suffixes of the $(\ell+1)$-mers in $\cL_{\ell+1}(\enc{cnst:naive}(\bfx))$. Then, by ordering and then removing the length-$\log(k)$ indices from these strings, we receive the string $\bfc = E(\bfx)$, and consequently, $\bfx$. Note that the role of the indices in this construction is crucial to deduce the string $\bfc = E(\bfx)$ from the set of strings $\widetilde{\bfc}_1,\widetilde{\bfc}_2,\ldots,\widetilde{\bfc}_k$. Without indices the order of these $k$ strings could have not been derived so we could only know the string $\bfc = E(\bfx)$ up to a permutation of its $k$ sub-strings. The next theorem analyzes the parameters of codes that can be constructed using \cref{cnst:naive} based upon the repeat-free encoders from~\cite{EliGabMedYaa21}.
\begin{theorem}\label{thm:red-naive}
Let $n,k$ grow such that $\alpha= \limsup\frac{\log(k)}{n}<1$. For the values of $\ell$ to be specified, take $m$ such that $\enc{cnst:naive}\colon \Sigma^m\to A_{n,k,\ell+1}$, and denote $\cC_A\deq \rng(\enc{cnst:naive})$. Then 
\begin{enumerate}
    \item One may choose a regime satisfying $\ell=\log(n k^2)+2(\log\log(n k)) + O(1)$, while assuring $\red(\cC_A) \leq \frac{q}{q-2}\cdot \frac{n k}{\log^2(n k)} + k \log(e) + o(k)$. (When $q=2$, this is $2q \frac{n k}{\log^2(n k)} + k \log(e) + o(k)$.) 
    \item Allowing $\ell = (1+\epsilon)\log(n k) + \log(k) + O(1)$ for any $\epsilon>0$, we have $\red(\cC_A) \leq 
    \frac{q(1-\alpha+o(1))^{1-\epsilon}}{q-2}\cdot (n k)^{1-\epsilon} 
    + k \log(e) + o(k)$. 
    (When $q=2$, this is $2q(1-\alpha+o(1))^{1-\epsilon}\cdot 
    (n k)^{1-\epsilon} + k \log(e) + o(k)$.)
    \item Finally, when $\epsilon>1$ in the definition of the last 
    part, we have that $\red(\cC_A) = k \log(e) + o(k)$.
\end{enumerate}
\end{theorem}

\vspace*{-0.5\baselineskip}
Note that the resulting redundancies are $o(nk)$.
\begin{IEEEproof}
\begin{enumerate}
\item
We may set, in the parameters of \cref{cnst:naive}, $\ell' = \log(n') + 
2\log\log(n') + 5$. We then have 
\begin{align*}
    \ell &= \ell'+\log(k) = \log(n' k) + 2\log\log(n') + 5 \\
    &= \log((n-\log(k))k^2) + 2\log\log((n-\log(k))k) + 5 \\
    &= \log(n k^2) + 2\log\log(nk) + O(1),
\end{align*}
where the last equality relies on $\alpha<1$.

We now utilize in \cref{cnst:naive} the repeat-free encoder described 
in \cite[Alg.~3]{EliGabMedYaa21}, which produces strings in  
$\rf_{\log(n')+2\log\log(n')+5}^{n'}$. It initializes by encoding an 
information string in $\Sigma^m$ into $Z(n',2\log\log(n'))$; this is done so that $0^{2\log\log(n')}$ may be used as a marker later on. 
Following steps of encoding the result into repeat-free strings 
interestingly require no further redundancy. We observed that 
efficient encoders exist into $Z(n',2\log\log(n'))$ using less than 
$\frac{q}{q-2}\cdot \frac{n'}{\log^2(n')}$ redundant symbols (in the 
case of $q=2$, the coefficient is based on \cite{LevYaa19} as 
described above).

It follows that $m = n' - \frac{q}{q-2}\cdot \frac{n'}{\log^2(n')}$, 
hence 
\end{enumerate}

\vspace{-1\baselineskip}
\begin{align*}
    \red(\cC_A) &= \log\abs*{\cX_{n,k}} - m \\
    &= \frac{q}{q-2} \frac{n k - k\log(k)}{\parenv*{\log(n k) + 
    O(1)}^2}  + k \log(e) + o(k) \\
    &\leq \frac{q}{q-2} \frac{n k}{\log^2(n k)} + k \log(e) + o(k).
\end{align*}

\begin{enumerate}\addtocounter{enumi}{1}
\item\label{par:red-naive-2}
Note that no part of the encoder of \cite[Alg.~3]{EliGabMedYaa21} is 
affected if initialization is done by encoding into 
$Z(n',\epsilon\log(n'))$ (and makers changed accordingly). It then 
produces strings in $\rf_{(1+\epsilon)\log(n')+5}^{n'}$. As observed, 
the initial encoding requires less than $\frac{q}{q-2}\cdot 
n'^{1-\epsilon}$ redundant symbols (and similarly for $q=2$). It 
follows that 
\begin{align*}
    \red(\cC_A) &= \log\abs*{\cX_{n,k}} - m \\
    &= \tfrac{q}{q-2} \left((n-\log(k))k\right)^{1-\epsilon} + k \log(e) + o(k) \\
    &\leq \tfrac{q(1-\alpha+o(1))^{1-\epsilon}}{q-2} 
    (n k)^{1-\epsilon} + k \log(e) + o(k).\vspace{-1ex}
\end{align*}
In this case, \vspace{-1ex}
\begin{align*}
    \ell &= \ell'+\log(k) = (1+\epsilon)\log(n') + \log(k) + 5 \\
    &= (1+\epsilon)\log(n k) + \log(k) + O(1).\vspace{-1ex}
\end{align*}

\item 
We analyze the cases where \cref{cnst:naive} may utilize the 
repeat-free encoder described in \cite[Alg.~1]{EliGabMedYaa21}, which 
produces strings in $\rf_{2\log(n')+2}^{n'}$ and requires a single 
redundant symbol (see \cite[Thm.~16]{EliGabMedYaa21}).

Note that $\ell' \geq 2\log(n') + 2$ if and only if 
\begin{align*}
    1+\epsilon &\geq \frac{2\log((n-\log(k))k) + 2}{\log(nk)} \\
    &= 2(1-o(1)) + o(1), 
\end{align*}
hence for sufficiently large $n,k$ and all $\epsilon>1$, the encoder 
of \cite[Alg.~1]{EliGabMedYaa21} may be used; in this case, $m=n'-1$, 
and 
\begin{align*}
    \red(\cC_A) &= \log\abs*{\cX_{n,k}} - m = k \log(e) + o(k).
\end{align*}
\end{enumerate}
\end{IEEEproof}

\begin{table*}[t]
	\caption{Redundancy and window length trade-off comparison}
	\begin{center}		
		\label{tab:compare}	\begin{tabular}{ccc}\hline
			Lower bound & \multicolumn{2}{c}{$\log(n k)-\ell = \omega_{n k}(1) \implies R(A_{n,k,\ell}) \leq R(B_{n,k,\ell}) = o_{n k}(1)$} \\
			\hline\hline
			Case & \cref{cnst:naive} &  \cref{cnst:sagi}  \\
			\hline
			\multirow{2}{*}{1} & $ \ell = \log(n k) + \log(k) + 2\log\log(n k) +O(1)  $ & $ \ell = \log(n k) + 2\log\log(n k) + 5 $ \\
			& $ \red(\cC_A) \leq \frac{q}{q-2}\cdot \frac{n k}{\log^2(n k)} + k \log(e) + o(k) $ & $ \red(\cC_B)\leq \frac{q}{q-2}\cdot \frac{n k}{\log^2(n k)} + k \log(n) (1+o(1)) $ \\
			\hline
            \multirow{2}{*}{2 ($\epsilon > 0$)} & $ \ell = (1+\epsilon)\log(n k) + \log(k) + O(1) $ & $\ell = (1+\epsilon)\log(n k) + 5 $ \\
			& $ \red(\cC_A) \leq \frac{q(1-\alpha+o(1))^{1-\epsilon}}{q-2}\cdot (n k)^{1-\epsilon} + k \log(e) + o(k) $ & $ \red(\cC_B) \leq \frac{q}{q-2}\cdot (n k)^{1-\epsilon} + k \log\parenv*{n^{1+\epsilon} k^\epsilon} (1+o(1)) $ \\
			\hline
			 \multirow{2}{*}{3} & $ \ell = (1+\epsilon)\log(n k) + \log(k) + O(1) $; $\epsilon > 1$ & $\ell = (1+\epsilon)\log(n k) + 5 $; $\epsilon \geq 1$ \\
			& $ \red(\cC_A) = k \log(e) + o(k) $ & $ \red(\cC_B) = k \log(n^{1+\epsilon} k^{\epsilon}) (1+o(1)) $ \\
			\hline
		\end{tabular}\vspace{-2.5ex}
	\end{center}\vspace{-1.5ex}
\end{table*}

\subsection{Construction B}
While in \cref{cnst:naive} we added indices in order to overcome the lack of ordering when the string $\bfc =E(\bfx)$ is partitioned into $k$ strings, in Construction B we tackle this constraint differently. We again partition $\bfc =E(\bfx)$ to $k$ strings but with overlapping segments between consecutive strings. The overlapping segments will guarantee in decoding that, given the set of $k$ strings, there will be only one way to concatenate them into one long string. As opposed to \cref{cnst:naive}, this also guarantees that there is no need to increase the length of the read $\ell$-mers with respect to the one required by the repeat-free encoders. 
\begin{construction}\label{cnst:sagi}
Let $\bfx\in\Sigma^m$ be an arbitrary information string, and encode it into an $\ell$-repeat-free string $\bfc=E(\bfx)\in\Sigma^{n'}$ using any known repeat-free encoder. 
For $n,k$ such that $n' = n k - (k-1) \ell = (n-\ell) k + \ell$, define $k$ length-$n$ strings
$\bfc_1,\ldots,\bfc_k\in\Sigma^n$ by $\bfc_i = 
\parenv*{c_{i,1},\ldots,c_{i,n}}$, where 
\begin{align*}
    c_{i,j} &\deq c_{(i-1)(n-\ell)+j}; \quad j=0,\ldots, n-1.
\end{align*}
Then, \vspace{-0.5ex}
\begin{align*}
\enc{cnst:sagi}(\bfx) \deq 
\bracenv*{\mathset*{\bfc_i}{i=1,\ldots,k}}\in \cX_{n,k}.
\end{align*}
\end{construction}
The decoding correctness of the information string $\bfx$ in Construction~\ref{cnst:sagi} follows from the following simple observation.
\begin{lemma}\label{lem:overlap}
For all $\bfx\in\Sigma^m$ it holds that 
$\cL_{\ell+1}(E(\bfx)) = \cL_{\ell+1}\parenv*{\enc{cnst:sagi}(\bfx)}$.
\end{lemma}
\begin{IEEEproof}
Since $\bfc_i$ is a substring of $\bfc$ for all $i$, it follows that 
$\cL_{\ell+1}\parenv*{\enc{cnst:sagi}(\bfx)}\subseteq 
\cL_{\ell+1}(E(\bfx))$. For the other direction, note that for all $1\leq 
i<k$, $\bfc_i,\bfc_{i+1}$ are overlapping substrings of $\bfc$, 
with a common substring of length $\ell$; thus all 
$(\ell+1)$-substrings of $\bfc$ are also substrings of some $\bfc_i$.
\end{IEEEproof}
\cref{lem:overlap} immediately implies the next corollary.
\begin{corollary}
$\enc{cnst:sagi}(\bfx)\in A_{n,k,\ell+1}$ for all $\bfx\in\Sigma^m$.
\end{corollary}
\begin{IEEEproof}
According to \cref{lem:overlap} and since $E(\bfx)\in \rf_\ell^{n'}$, the corollary's statement follows.
\end{IEEEproof}

\begin{comment}
\begin{table*}[h!]
	\caption{\yy{Is this more readable?}}
	\begin{center}		
		\label{tab:compare}	\begin{tabular}{ccc}\hline
			Case & \cref{cnst:naive} &  Construction B  \\
			\hline
			$ \ell = \log(n k) + 2\log\log(n k) $ & $ + \log(k) +O(1) $ & $ + 5 $ \\
			$ \red(C) \leq \frac{q}{q-2}\cdot \frac{n k}{\log^2(n k)} $ & $ + k \parenv*{\log(e)+o_k(1)} $ & $ + k \log(n) (1+o(1)) $ \\
			\hline
            $\epsilon > 0$: $ \ell = (1+\epsilon)\log(n k) + O(1) $ & $ + \log(k) $ & $ + 0 $ \\
			$ \red(C) \leq \frac{q}{q-2}\cdot (n k)^{1-\epsilon} $ & $ \cdot (1-\alpha+o(1))^{1-\epsilon} + k \parenv*{\log(e)+o_k(1)} $ & $ + k \log\parenv*{n^{1+\epsilon} k^\epsilon} (1+o(1)) $ \\
			\hline
			 $ \ell = (1+\epsilon)\log(n k) + O(1) $ & $ + \log(k) $; $\epsilon > 1$ & $\epsilon = 1$ \\
			$ \red(C) = $ & $ k \parenv*{\log(e)+o_k(1)} $ & $ k \log(n^2 k) (1+o(1)) $ \\
			\hline
		\end{tabular}
	\end{center}
\end{table*}
\end{comment}
We are now ready to analyze the code parameters that \cref{cnst:sagi} can achieve, again using the repeat-free encoders from~\cite{EliGabMedYaa21}. 
\begin{theorem}\label{thm:red-sagi}
Let $n,k$ grow such that $\alpha = \limsup\frac{\log(k)}{n}<1$. For 
the values of $\ell$ to be specified, take $m$ such that 
$\enc{cnst:sagi}\colon \Sigma^m\to A_{n,k,\ell+1}$, and denote 
$\cC_B\deq \rng(\enc{cnst:sagi})$. 
Then 
\begin{enumerate}
    \item Letting $\ell = \log(n k) + 2\log\log(n k) + 5$, we have $\red(\cC_B)\leq \frac{q}{q-2}\cdot \frac{n k}{\log^2(n k)} + k \log(n) (1+o(1))$. (When $q=2$, this is $2q \frac{n k}{\log^2(n k)} + k \log(n) (1+o(1))$.)
    \item Setting $\ell = (1+\epsilon)\log(n k) + 5$ for any $\epsilon>0$, we have $\red(\cC_B) \leq \frac{q}{q-2}\cdot (n k)^{1-\epsilon} + k \log\parenv*{n^{1+\epsilon} k^\epsilon} (1+o(1))$. (For $q=2$, this is $2q (n k)^{1-\epsilon} + k \log\parenv*{n^{1+\epsilon} k^\epsilon} (1+o(1))$.)
    \item Finally, when $\epsilon\geq 1$ in the definition of the last part, we have $\red(\cC_B) = k \log(n^{1+\epsilon} k^{\epsilon}) (1+o(1))$.
\end{enumerate}
\end{theorem}
\begin{IEEEproof}
\begin{enumerate}
\item\label{par:red-sagi-1}
We start by observing, since $n'\leq n k$, that the repeat-free 
encoder of \cite[Alg.~3]{EliGabMedYaa21} may produce strings in 
$\rf_{\ell}^{n'}$, if initialization is done by encoding into 
$Z(n',2\log\log(n k))$ (instead of $Z(n k,2\log\log(n k))$). 
Following steps \cite[Lem.~19--20, Lem.~23]{EliGabMedYaa21} are 
not affected.

In this case, we have $m = n' - \frac{q}{q-2}\cdot 
\frac{n'}{\log^2(nk)}$ (again, the case $q=2$ is handled similarly). 
Therefore 
\end{enumerate}

\vspace{-1\baselineskip}
\begin{align*}
    \red(\cC_B) &= \log\abs*{\cX_{n,k}} - m \\
    &= \frac{q}{q-2}\cdot \frac{n k}{\log^2(n k)} + k (\ell - \log(k)) + O(k) - \ell \\
    &= \frac{q}{q-2}\cdot \frac{n k}{\log^2(n k)} + 
    k \log(n) (1+o(1)).
\end{align*}

\begin{enumerate}\addtocounter{enumi}{1}
\item\label{par:red-sagi-2}
The next part follows similarly to \cref{par:red-sagi-1}, based on 
the adjusted encoder described in \cref{par:red-naive-2} of 
\cref{thm:red-naive}.

\item\label{par:red-sagi-3}
Finally, note that is suffices that $\ell = (1+\epsilon)\log(n k) + 5 > 
2\log(n') + 2$ to utilize the repeat-free encoder of 
\cite[Alg.~1]{EliGabMedYaa21}; in this case, $m = n'-1$, hence 
\begin{align*}
    \red(\cC_B) &= \log\abs*{\cX_{n,k}} - m \\
    &= k (\ell - \log(k)) + k \log(e) + o(k) - \ell + 1 \\
    &= k \log(n^{1+\epsilon} k^{\epsilon}) (1+o(1)).\vspace{-1ex}
\end{align*}
\end{enumerate}

\vspace{-3ex}
\end{IEEEproof}

It should be noted that \cref{thm:red-sagi} does not preclude the 
possibility that the encoder of \cref{par:red-sagi-3} requires more 
redundancy than that of \cref{par:red-sagi-1} (namely, in any 
asymptotic regime satisfying $\log(k)=\Omega(n)$); this is an oddity 
since $A_{n,k,\ell_1}\subseteq A_{n,k,\ell_2}$ for all 
$\ell_1\leq\ell_2$. We observe that it is inherent to 
\cref{cnst:sagi} that the last step might introduce more redundancy 
than is required by $E$, the repeat-free encoder utilized. 
Nevertheless, \cref{thm:red-sagi} is structured to take into account 
other asymptotic regimes, and should be applied accordingly in 
practice.

\vspace{-1ex}
\subsection{Comparison and Summary}
We first seek to give a converse to \cref{lem:zero-rate} and establish the result on the minimum value of $\ell$ which guarantees that the asymptotic rate of multi-strand $\ell$-reconstruction codes (in fact, $R(A_{n,k,\ell})$) is $1$. This result is established in the next corollary using the results of \cref{cnst:sagi}. 

\vspace*{-0.5\baselineskip}
\begin{corollary}\label{cor:rate-one}
For sufficiently large $n,k$ satisfying $\limsup\frac{\log(k)}{n}<1$ and for $ \ell \ge \log(n k) + 2\log\log(n k) + 5 $, it holds that  $R(B_{n,k,\ell}) \geq R(A_{n,k,\ell}) = 1-o_{nk}(1). $
\end{corollary}

\vspace*{-0.5\baselineskip}
\begin{IEEEproof}
Observe that under the assumption $n-\log(k)=\Omega(n)$ we have 
from \cref{lem:channel} that $\log\abs*{\cX_{n,k}}=\Omega(n k)$. The 
proposition of \cref{par:red-sagi-1} of \cref{thm:red-sagi} now 
suffices to establish the corollary's statement.
\end{IEEEproof}

Note that if one aims to achieve the same result using \cref{cnst:naive}, then the minimum value of $\ell$ should be $\log(n k) + \log(k) + 2\log\log(n k) +O(1)$ and hence there is a gap of roughly $\log(k)$ with respect to the result in \cref{cor:rate-one}. However, \cref{cnst:naive} can support better redundancy for comparable values of $\ell$. Since \cref{cnst:naive,cnst:sagi} provide codes with parameters that cannot be directly compared, \cref{tab:compare} lists the parameters of each construction for different regimes of $\ell$ in order to have a better understanding of the trade-offs between the minimal window length $\ell$ and the constructions' redundancy from \cref{thm:red-naive,thm:red-sagi}. In general, one can observe that the window length in \cref{cnst:naive} should be larger than the one in \cref{cnst:sagi} but construction's redundancy is smaller.

Before concluding, we suggest that one might consider a slightly 
different channel definition to the one handled above where the $k$ strands are 
required to be distinct from one another, i.e., when information is 
stored in the space $ \cX^*_{n,k} \deq \mathset*{S\subseteq \Sigma^n}{\abs*{S}=k}$.
A priori, it seems feasible that the added restriction might allow 
for lower redundancy (when measured in $\cX^*_{n,k}$). However, we note 
that $\abs*{\cX^*_{n,k}} = \binom{q^n}{k}$, thus a similar development 
to \cref{lem:channel} yields \vspace{-1ex}
\begin{align*}
    q^{nk} \parenv*{\frac{e}{k} - \frac{e}{2q^n}}^k 
    \leq \abs*{\cX^*_{n,k}} 
    \leq \frac{q^{nk}}{k!}.
\end{align*}
It follows that $\log\abs*{\cX^*_{n,k}} = n k - k \log(k) + 
k \log(e) + o(k)$ as well. A careful examination reveals 
that \cref{cnst:naive,cnst:sagi} actually encode into the set 

\vspace*{-1.25\baselineskip}
\begin{align*}
    \mathset*{S\in \cX^*_{n,k}}{S\ \text{has a unique $\ell$-profile}},
\end{align*}
and hence the results of this work also hold for this setup of the problem.


\begin{thebibliography}{10}
\providecommand{\url}[1]{#1}
\csname url@samestyle\endcsname
\providecommand{\newblock}{\relax}
\providecommand{\bibinfo}[2]{#2}
\providecommand{\BIBentrySTDinterwordspacing}{\spaceskip=0pt\relax}
\providecommand{\BIBentryALTinterwordstretchfactor}{4}
\providecommand{\BIBentryALTinterwordspacing}{\spaceskip=\fontdimen2\font plus
\BIBentryALTinterwordstretchfactor\fontdimen3\font minus
  \fontdimen4\font\relax}
\providecommand{\BIBforeignlanguage}[2]{{%
\expandafter\ifx\csname l@#1\endcsname\relax
\typeout{** WARNING: IEEEtranS.bst: No hyphenation pattern has been}%
\typeout{** loaded for the language `#1'. Using the pattern for}%
\typeout{** the default language instead.}%
\else
\language=\csname l@#1\endcsname
\fi
#2}}
\providecommand{\BIBdecl}{\relax}
\BIBdecl

\bibitem{AchDasMilOrlPan10}
J.~Acharya, H.~Das, O.~Milenkovic, A.~Orlitsky, and S.~Pan, ``On reconstructing
  a string from its substring compositions,'' in \emph{Proc. of the IEEE
  International Symposium on Information Theory}, Austin, Texas, USA, 2010, pp.
  1238--1242.

\bibitem{AchDasMilOrlPan15}
------, ``String reconstruction from substring compositions,'' \emph{SIAM
  Journal on Discrete Mathematics}, vol.~29, no.~3, pp. 1340--1371, 2015.

\bibitem{BatKanKhaMcG2004}
T.~Batu, S.~Kannan, S.~Khanna, and A.~McGregor, ``Reconstructing strings from
  random traces,'' in \emph{Proc. of the Fifteenth Annual ACM-SIAM Symposium on
  Discrete Algorithms, New Orleans, USA}, 2004, pp. 910--918.

\bibitem{BreBreTse13}
G.~Bresler, M.~Bresler, and D.~Tse, ``Optimal assembly for high throughput
  shotgun sequencing,'' \emph{{{BMC}} Bioinformatics}, vol.~14, 2013.

\bibitem{ChaChrEzeKia17}
Z.~{Chang}, J.~{Chrisnata}, M.~F. {Ezerman}, and H.~M. {Kiah}, ``Rates of {DNA}
  sequence profiles for practical values of read lengths,'' \emph{IEEE
  Trans.~on Inform.~Theory}, vol.~63, no.~11, pp. 7166--7177, Nov. 2017.

\bibitem{Chin_2013}
C.-S. Chin, D.~H. Alexander, P.~Marks, A.~A. Klammer, J.~Drake, C.~Heiner,
  A.~Clum, A.~Copeland, J.~Huddleston, E.~E. Eichler, S.~W. Turner, and
  J.~Korlach, ``Nonhybrid, finished microbial genome assemblies from long-read
  {SMRT} sequencing data,'' \emph{Nature Methods}, vol.~10, no.~6, pp.
  563--569, 2013.

\bibitem{DudSch2003}
M.~Dud{\i}k and L.~J. Schulman, ``Reconstruction from subsequences,''
  \emph{Journal of Combinatorial Theory, Series A}, vol. 103, no.~2, pp.
  337--348, 2003.

\bibitem{EliGabMedYaa21}
O.~{Elishco}, R.~{Gabrys}, M.~{M\'{e}dard}, and E.~{Yaakobi}, ``Repeat-free
  codes,'' \emph{IEEE Trans.~on Inform.~Theory}, 2021.

\bibitem{GabMil18}
R.~Gabrys and O.~Milenkovic, ``Unique reconstruction of coded sequences from
  multiset substring spectra,'' in \emph{Proc. of the IEEE International
  Symposium on Information Theory, Vail, Colorado, USA}, Jun. 2018, pp.
  2540--2544.

\bibitem{GabPatMil21}
R.~{Gabrys}, S.~{Pattabiraman}, and O.~{Milenkovic}, ``Reconstructing mixtures
  of coded strings from prefix and suffix compositions,'' in \emph{Proceedings
  of the 2020 {IEEE} Information Theory Workshop ({ITW}'2020), Riva del Garda,
  Italy}, Apr. 2021.

\bibitem{GaMoRa16}
S.~Ganguly, E.~Mossel, and M.~Racz, ``Sequence assembly from corrupted shotgun
  reads,'' in \emph{Proc. of the IEEE International Symposium of Information
  Theory}, Barcelona, Spain, 2016, pp. 265--269.

\bibitem{JaiFarSchBru17a}
S.~Jain, F.~Farnoud, M.~Schwartz, and J.~Bruck, ``Duplication-correcting codes
  for data storage in the {DNA} of living organisms,'' \emph{IEEE Trans.~on
  Inform.~Theory}, vol.~63, no.~8, pp. 4996--5010, Aug. 2017.

\bibitem{KhaPerBabNavSho18}
A.~R. {Khan}, M.~T. {Pervez}, M.~E. {Babar}, N.~{Naveed}, and M.~{Shoaib}, ``A
  comprehensive study of de novo genome assemblers: Current challenges and
  future prospective,'' \emph{Evolutionary Bioinformatics}, vol.~14, Jan. 2018,
  {PMID}: 29511353.

\bibitem{Lev01}
V.~I. Levenshtein, ``Efficient reconstruction of sequences from their
  subsequences or supersequences,'' \emph{Journal of Combinatorial Theory,
  Series A}, vol.~93, no.~2, pp. 310--332, 2001.

\bibitem{LevYaa19}
M.~{Levy} and E.~{Yaakobi}, ``Mutually uncorrelated codes for {{DNA}}
  storage,'' \emph{IEEE Trans.~on Inform.~Theory}, vol.~65, no.~6, pp.
  3671--3691, Jun. 2019.

\bibitem{LoNiQuJoSiJa15}
N.~Loman, J.~Quick, and J.~Simpson, ``A complete bacterial genome assembled de
  novo using only nanopore sequencing data,'' \emph{Nature Methods}, vol.~12,
  no.~8, pp. 733--–735, 2015.

\bibitem{MacSlo78}
F.~J. MacWilliams and N.~J.~A. Sloane, \emph{The Theory of Error-Correcting
  Codes}.\hskip 1em plus 0.5em minus 0.4em\relax North-Holland, 1978.

\bibitem{BenMeySchSmiSto91}
B.~Manvel, A.~Meyerowitz, A.~Schwenk, K.~Smith, and P.~Stockmeyer,
  ``Reconstruction of sequences,'' \emph{Discrete Mathematics}, vol.~94, no.~3,
  pp. 209--219, 1991.

\bibitem{MarYaa21}
S.~{Marcovich} and E.~{Yaakobi}, ``Reconstruction of strings from their
  substrings spectrum,'' \emph{IEEE Trans.~on Inform.~Theory}, vol.~67, no.~7,
  pp. 4369--4384, Jul. 2021.

\bibitem{MarRotSie01}
\BIBentryALTinterwordspacing
B.~H. Marcus, R.~M. Roth, and P.~H. Siegel, ``An introduction to coding for
  constrained systems,'' Oct. 2001, unpublished Lecture Notes. [Online].
  Available: \url{www.math.ubc.ca/~marcus/Handbook}
\BIBentrySTDinterwordspacing

\bibitem{MotBreTse13}
A.~S. Motahari, G.~Bresler, and D.~Tse, ``Information theory of {{DNA}} shotgun
  sequencing,'' \emph{IEEE Transactions on Information Theory}, vol.~59,
  no.~10, pp. 6273--6289, 2013.

\bibitem{MoRaTs13}
A.~Motahari, K.~Ramchandran, D.~Tse, and N.~Ma, ``Optimal {{DNA}} shotgun
  sequencing: Noisy reads are as good as noiseless reads,'' in \emph{Proc. of
  the IEEE International Symposium of Information Theory}, Istanbul, Turkey,
  2013, pp. 1640--1644.

\bibitem{salzberg_2010}
S.~L. Salzberg, ``Mind the gaps,'' \emph{Nature Methods}, vol.~7, no.~2, pp.
  105--106, 2010.

\bibitem{Sco97}
A.~D. Scott, ``Reconstructing sequences,'' \emph{Discrete Mathematics}, vol.
  175, no. 1-3, pp. 231--238, 1997.

\bibitem{SCT15}
I.~{Shomorony}, T.~{Courtade}, and D.~{Tse}, ``Do read errors matter for genome
  assembly?'' in \emph{Proc. of the IEEE International Symposium of Information
  Theory}, Hong Kong, 2015, pp. 919--923.

\bibitem{ShKaXiCoTs16}
I.~Shomorony, G.~Kamath, F.~Xia, T.~Courtade, and D.~Tse, ``Partial {{DNA}}
  assembly: A rate-distortion perspective,'' in \emph{Proc. of the IEEE
  International Symposium of Information Theory}, Barcelona, Spain, 2016, pp.
  1799--1803.

\bibitem{Ukk1992}
E.~Ukkonen, ``Approximate string-matching with q-grams and maximal matches,''
  \emph{Theoretical Computer Science}, vol.~92, no.~1, pp. 191--211, 1992.

\bibitem{YehPol21}
Y.~{Yehezkeally} and N.~{Polyanskii}, ``On codes for the noisy substring
  channel,'' in \emph{Proceedings of the 2021 {IEEE} International Symposium on
  Information Theory ({ISIT}'2021), Melbourne, Victoria, Australia}, Jul. 2021.

\end{thebibliography}
\end{document}